\documentclass[prd,nofootinbib,showpacs,12pt]{revtex4-1}
\DeclareRobustCommand{\baselinestretch{1.3}}

\usepackage{amsmath,amssymb,amsthm}

\def\bk{{\bf k}}
\def\x{{\bf x}}
\def\y{{\bf y}}
\def\n{{\bf n}}
\def\im{{\rm i}}
\def\P{{\rm P}}
\def\B{{\rm B}}

\begin{document}

\title{Statistical anisotropy as a consequence of inflation\footnote{Grassmannian
Conference in Fundamental Cosmology, September 14--19, 2009, Szczecin, Poland}}

\author{Yuri Shtanov} \email{shtanov@bitp.kiev.ua}
\affiliation{Bogolyubov Institute for Theoretical Physics, Kiev 03680, Ukraine} %

\begin{abstract}
Cosmological inflation remains to be a unique mechanism of generation of plausible
initial conditions in the early universe.  In particular, it generates the primordial
quasiclassical perturbations with power spectrum determined by the fundamental principles
of quantum field theory.  In this work, we pay attention to the fact that the
quasiclassical perturbations permanently generated at early stages of inflation break
homogeneity and isotropy of the cosmological background. The evolution of the small-scale
quantum vacuum modes on this inhomogeneous background results in statistical anisotropy
of the primordial power spectrum, which can manifest itself in the observable large-scale
structure and cosmic microwave background. The effect is predicted to have almost
scale-invariant form dominated by a quadrupole and may serve as a non-trivial test of the
inflationary scenario.  Theoretical expectation of the magnitude of this statistical
anisotropy depends on the assumptions about the physics in the trans-Planckian region of
wavenumbers.
\end{abstract}


\maketitle

Cosmological inflation remains to be the only realistic scenario that addresses several
of the long-standing puzzles of the standard cosmology and also explains the origin of
primordial density perturbations \cite{Linde}. The predictions of the inflationary theory
--- most notably, the spatial flatness of the universe and adiabatic primordial
perturbations with almost scale-invariant spectrum --- are surprisingly well verified by
modern observations of the large-scale structure (LSS) and cosmic microwave background
(CMB).  In view of this success, it is rather important and intriguing to be able to test
this scenario further.

The role of quantum fluctuations on the inflationary stage was revealed in several
aspects. Primordial perturbations for the formation of the observable structure are
generated at the late stages of inflation (the last 60 or so $e$-foldings). However,
depending on a concrete realization, the inflationary stage can be rather long, in
principle, starting right from the Planck energy density. The total number of
inflationary $e$-foldings can be enormous, with typical values $\sim 10^{12}$ in the
scenario of chaotic inflation. It is clear that generation of super-Hubble scalar and
tensor perturbations with quasiclassical features cannot be considered as specific to the
last stage of inflation, but that this process should occur all along the inflationary
history. This observation is the basis for the theory of the self-regenerating eternal
inflationary universe (see \cite{Linde}). According to this picture, we are living today
in a relatively small domain in which inflation has ended some thirteen billion years
ago, while the universe as a whole is very inhomogeneous, with its remote parts still
experiencing inflation.

A question arises whether it is possible to verify this picture of the universe.
Obviously, a positive answer to this question would imply the existence of information in
our observable realm about the structure of the universe on extremely large scales. At
first sight, this possibility is excluded.  Indeed, photons or neutrino cannot carry such
an information: they simply do not exist at the inflationary stage. The primordial
gravitational waves could, in principle, be a good candidate, but they remain to be
undetected.

In our work \cite{CS,SP}, we proposed an idea that the large-scale distribution of
galaxies and clusters can contain specific information about the inflationary epoch.
According to the inflationary scenario, before becoming the observable large-scale
structure, the inhomogeneities on the specified comoving spatial scales at the
inflationary stage existed in the form of quantum fluctuations.  Propagating with the
speed close to the speed of light, they passed over enormous comoving distances and thus
potentially can carry information about the inhomogeneity of the inflationary universe.
This bears some similarity with the Sachs--Wolfe effect \cite{SW}, in which the
large-scale inhomogeneity of the universe at a later cosmological epoch is responsible
for the locally observed anisotropy in the temperature of the cosmic microwave background
radiation.

Technically, the idea is to take into account the long-wave (super-Hubble) {\em
quasiclassical inhomogeneity\/} of the inflationary universe into account when
considering the evolution of small-scale (sub-Hubble) quantum fluctuations. The
background inhomogeneities generated during inflation are of two types: the scalar ones,
described by the relativistic potential $\Phi$, and the tensor ones, described by the
transverse traceless metric perturbations $h_{ij}$, so that the background metric during
inflation can be written as
\begin{equation} \label{background}
ds^2 = a^2 (\eta) \left[ \left(1 + 2 \Phi_\B \right) d \eta^2 - \left(1 - 2 \Phi_\B \right) d \x^2
+ h_{ij}^\B dx^i dx^j \right] \, ,
\end{equation}
where $\Phi_\B$ and $h_{ij}^\B$ are the background parts of perturbations, on scales
exceeding the Hubble scales during inflation.  These (large-scale) perturbations can be
considered as quasiclassical parts of the corresponding fields \cite{Linde}.  The idea is
then to study the behavior of the small-scale modes that determine the vacuum state on
the background of (\ref{background}).  On small scales, one can neglect the effects of
self-gravity, and the problem reduces to solving the wave equation for the perturbations
$\delta \varphi$ of the inflaton field $\varphi$ (which we take to be a free field with
mass $m_\varphi$):
\begin{equation}
\Box_\B \delta\varphi + m_\varphi^2 \delta\varphi = 0 \, ,
\end{equation}
where $\Box_\B$ is the wave operator for the background metric (\ref{background}).

The presence of the background inhomogeneities $\Phi_\B$ and $h_{ij}^\B$ leads to the
appearance of additional phases in the modes $\delta\varphi_\bk$ and $\Phi_\bk$, which
``distort'' their spatial form from the original plane waves $\propto e^{\im \bk \x}$.
As a result, the primordial power spectrum of scalar perturbations acquires {\em
statistical anisotropy\/} of quite specific form. Namely, calculation of the correlation
function of the relativistic potential $\Phi$ at the end of inflation gives \cite{CS,SP}
\begin{equation} \label{cs-1}
\langle \Phi (\x) \Phi (\y) \rangle = \int \frac{d^3 \bk}{(2 \pi k)^3} P_k
\left( 1 +  \nu_\bk \right)
e^{\im \bk (\x - \y)}  \, ,
\end{equation}
where $P_k$ describes the standard isotropic power spectrum, and $\nu_\bk$ is a new
anisotropic correction, which turns out to be almost scale-invariant. Specifically, its
multipole expansion is given by
\begin{equation} \label{cs-2}
\nu_\bk = \left(n_S - 1 \right) \left[ \Lambda_{ij}\, n^i n^j +
\Lambda_{ijkl}\, n^i n^j n^k n^l + \ldots \right] \, ,
\end{equation}
where $n_S (k)$ is the spectral index of scalar perturbations (which is known to be close
to unity), and the quantities $\Lambda_{\cdots} (k)$ are effectively traceless and depend
on $k$ very weakly. Moreover, expression (\ref{cs-2}) is essentially dominated by the
first term, with quadrupole dependence on the unit vector $n^i = k^i / k$.

The expected magnitude of the anisotropic part (\ref{cs-2}) is given by the cosmic
variance
\begin{equation} \label{variance}
\left\langle \nu^2_\bk \right\rangle \simeq
(n_S - 1)^2 \frac{G \left( H_\im^4 - H_k^4 \right)}{10 \pi m_\varphi^2} \approx
2 \times 10^{-3}\, G^3 \left( V_\im \varphi_\im^2 - V_k \varphi_k^2 \right) \, ,
\end{equation}
where the index ``$k$'' in the homogeneous quantities refers to the moment of the
Hubble-radius crossing by the wave with comoving wavenumber $k$. Expression
(\ref{variance}) crucially depends on the moment of time where one sets the initial
(vacuum) conditions for the quantum fluctuations of interest at the inflationary state
(this is denoted by the index ``i''). The earlier this moment is, the larger is the
number of inhomogeneous quasiclassical modes that have affected the evolution of the
small-scale quantum fluctuations, hence, the larger is the expected statistical
anisotropy.

The result (\ref{variance}) gives rise to an important issue that was first discussed in
\cite{CS} and then elaborated in \cite{SP}. It is observed that the physical wave number
or frequency of a particular quantum mode in the cosmological frame exceeds the Planck
value $M_\P \simeq G^{-1/2}$ during most part of the inflationary universe. A natural
question arises, whether it is legitimate to use the standard field theory in cosmology
on spatial scales below the Planck length. If the trans-Planckian issue can be safely
ignored, then the statistical anisotropy can be significant, with the dominant term
$\Lambda_{ij}$ in (\ref{cs-2}) being up to the order of unity (this is where our linear
approximation in $h_{ij}^\B$ breaks down). In this case, we get $\left\langle \nu_\bk^2
\right\rangle \simeq 2 \times 10^{-3}$.  If, however, one is allowed to specify locally
homogeneous and isotropic initial conditions for quantum modes only after the
Planck-radius crossing, then the resulting statistical anisotropy is rather small,
$\left\langle \nu_\bk^2 \right\rangle \sim 10^{-14}$.

There are several arguments in favor of the possibility of working in the trans-Planckian
frequency region.  First of all, one can note that the wavelength and frequency of a
particular mode are not Lorentz-invariant quantities.  Thus, the authors of
\cite{tp-creation} argue that nothing dangerous results from this procedure {\em in a
locally Lorentz-invariant theory\/} as long as the invariant $k^\mu k_\mu \ll M_\P^2$.
Similar arguments in cosmological context are given in \cite{VK}, and the authors of
\cite{AC} consider the trans-Planckian effects in quantum gravity and argue in favor of
the asymptotic safety in this region. One can also note that quite a similar
trans-Planckian problem arises in the theory of quantum radiation from a black hole which
is formed as a result of gravitational collapse (see \cite{bh} for a review). While the
prediction of quantum radiation of black holes hardly can be tested experimentally for
black holes with realistic masses, we have a different situation in the case of the
cosmological prediction under consideration in this paper, which, in principle, can be
verified by observations. In what follows, we discuss such testable predictions.

Statistical anisotropy of the power spectrum of primordial scalar perturbations will be
manifest in the LSS and CMB.  The anisotropic factor $\nu_\bk$ is inherited in the power
spectrum of the density contrast $\delta = \delta \rho / \rho$.  The corresponding
correlation function is then the sum of isotropic and anisotropic parts:
\begin{equation} \label{lss}
\xi (\x) = \xi_0 (x) + \xi_1 (\x) \, , \quad
\xi_0 (x) = \frac{1}{2 \pi^2} \int dk k^2 \delta_k^2 \frac{\sin kx}{kx} \, , \quad
\xi_1 (\x) = \int \frac{d^3 \bk}{(2 \pi)^3} \delta_k^2 \nu_\bk e^{\im \bk \x} \, ,
\end{equation}
in which usual infrared and ultraviolet cut-offs are implied in the integration over wave
vectors.

The anisotropic quantity $\nu_\bk$ is dominated by the first term in (\ref{cs-2}). In
view of the fact that the product $ \left( n_S - 1 \right) \Lambda_{ij}$ depends very
weakly on the wave number $k$, one can treat it as a constant and take out of the
integral in (\ref{lss}). Since the matrix $\Lambda_{ij}$ is traceless, one gets the
following expression:
\begin{equation} \label{xi1}
\xi_1 (\x) \approx \left( n_S - 1 \right) \Lambda_{ij} \, f(x) \frac{x^i x^j}{x^2} \, ,
\quad f(x) = \xi_0 (x) + \frac{3}{2 \pi^2} \int dk k^2 \delta_k^2 \frac{ kx \cos kx - \sin kx}{(kx)^3}
\, .
\end{equation}

The specific form (\ref{xi1}) of the anisotropic part of the correlation function, in
principle, can be tested by observations. Measurement of the quadrupole of the power
spectrum can simply be done by comparing the amplitudes of Fourier modes in different
directions. The standard error to the quadrupole coefficients $q_{ij} = \left( n_S - 1
\right) \Lambda_{ij}$ using the SDSS data is estimated in \cite{PK} to be $\sigma_q \sim
10^{-2}$.

Statistical anisotropy of the primordial power spectrum in the form (\ref{cs-1}),
(\ref{cs-2}) belongs to the category that was recently studied in detail in \cite{PK}. It
is given by the general formula
\begin{equation} \label{PK1}
P_\bk = P_k \left[ 1 + \sum_{LM} g_{LM} (k) Y_{LM} (\n) \right] \, ,
\end{equation}
where $P_k$ is the isotropic part of the anisotropic power spectrum $P_\bk$, and $g_{LM}
(k)$ are the coefficients of the expansion of the statistical anisotropy into spherical
harmonics $Y_{LM} (\n)$, where $\n = \bk / k$.  In our case, the sum in (\ref{PK1}) is
dominated by harmonics with $L = 2$, with coefficients $g_{2M} \sim \left( n_S - 1
\right) \Lambda_{ij}$ that depend on $k$ very weakly.

As usual, one can expand the CMB temperature map $T (\hat \n)$, depending on a direction
on the sky $\hat \n$, into spherical harmonics and calculate the covariance matrix:
\begin{equation}
T(\hat \n) = T_0 \sum_{lm} a_{lm} Y_{lm} (\hat \n)\, , \qquad
\left\langle a^{}_{lm} a^*_{l'm'} \right\rangle = \delta_{ll'} \delta_{mm'} C_l +
\sum_{LM} \xi^{LM}_{lml'm'} D^{LM}_{ll'} \, ,
\end{equation}
with
\begin{eqnarray}
&&C_l = \frac2\pi \int_0^\infty dk k^2 P_k \Theta_l^2 (k) \, , \qquad
\xi^{LM}_{lml'm'} = \int d \hat \n \, Y_{lm}^* (\hat \n) Y_{l'm'}^{} (\hat \n)
Y_{LM}^{} (\hat \n) \, , \\
&&D^{LM}_{ll'} = \frac{2 (-\im)^{l - l'}}{\pi} \int_0^\infty dk k^2 P_k g_{LM} (k)
\Theta_l (k) \Theta_{l'} (k) \, .
\end{eqnarray}
Here, $\Theta_l(k)$ is the kernel of the linear relation between $T (\hat \n)$ and
$\delta_\bk$ describing the contribution to the $l$th temperature moment from the wave
vector $\bk$ [with these conventions, $\Theta_l(k)$ is a real function].

The quantities $D^{LM}_{ll'}$ are the generalization of the coefficients $C_l$ to the
case of statistical anisotropy.  In the case where the functions $g_{LM} (k)$ can be
treated as constants, there arises the relation $D^{LM}_{ll'} = g_{LM} W_{ll'} $ with
$W_{ll'}$ depending only on the isotropic cosmology (in particular, $W_{ll} = C_l$).
Thus, the combinations $D^{LM}_{ll'} / W_{ll'}$ can be regarded as convenient estimators
for the coefficients $g_{LM}$ in this case.  The estimators for $D^{LM}_{ll'}$ were
constructed in \cite{PK} and essentially coincide with the bipolar-spherical-harmonic
coefficients of \cite{HS}. For the case $L = 2$ of interest for our work, it is shown in
\cite{PK} that the variances with which the quantities $g_{2M}$ can be measured using the
temperature maps of WMAP and Planck are $\sigma_{g} \sim 1.2 \times 10^{-2}$ and $3.8
\times 10^{-3}$, respectively.  This is comparable to the highest theoretical estimates
that we get within the limits of our approximation, linear in the background
inhomogeneity of the inflationary universe.  In a complete nonlinear calculation,
therefore, the resulting statistical anisotropy may well exceed the current threshold of
its detection.

In conclusion, we note that our work represents a theoretical contribution to the issue
of the possible origin of statistical anisotropy in the universe. The predictions made in
this paper have an advantage of being based on the usual theory of chaotic inflation
without any additional physical ingredients. Observation of statistical anisotropy in the
scale-invariant form (\ref{cs-2}) may thus be regarded as a nontrivial confirmation of
the chaotic inflation scenario; this is the only possible test of chaotic inflation that
we are aware of.  On the other hand, since the magnitude of the statistical anisotropy
depends on both the details of the inflationary scenario and the physical assumptions
about the trans-Planckian region, its nondetection by itself will not invalidate any of
the existing models of inflation.

\section*{Acknowledgments}

This contribution, as well as participation of the author in the Grassmannian Conference
in Fundamental Cosmology in Szczecin, Poland, was supported by the J{\'o}sef Mianowski
Fund and Foundation for Polish Science.  The work was also supported in part by the
``Cosmomicrophysics'' programme and Program of Fundamental Research of the Physics and
Astronomy Division of the National Academy of Sciences of Ukraine, and by the State
Foundation for Fundamental Research of Ukraine under grant F28.2/083.

\end{document}